\documentclass[11pt]{article}

\usepackage[utf8]{inputenc}
\usepackage[a4paper, margin=1in]{geometry}

\usepackage{amssymb}
\usepackage{amsmath}
\usepackage{bm}
\usepackage{mathtools}
\usepackage{adjustbox}
\usepackage[bibliography=common]{apxproof}
\newtheorem{theorem}{Theorem}
\newtheorem{definition}[theorem]{Definition}
\newtheoremrep{theorem1}[theorem]{Theorem}
\newtheoremrep{lemma}[theorem]{Lemma}

\renewcommand{\vec}[1]{\bm{#1}} 

\usepackage{tikz}
\usetikzlibrary{positioning}
\usetikzlibrary{arrows}
\usetikzlibrary{arrows.meta}
\usepackage{pgfplots}
\pgfplotsset{compat=1.15}
\usepackage{subcaption}
\usepackage{xcolor}
\definecolor{myGray}{rgb}{0.65,0.65,0.65}

\DeclarePairedDelimiter\abs{\lvert}{\rvert}%
\renewcommand{\epsilon}{\varepsilon}
\renewcommand{\phi}{\varphi}

\usepackage{float}

\usepackage[
backend=biber,
style=alphabetic, 
sorting=ynt
]{biblatex}
\usepackage{authblk}
\begin{document}

\title{Bailouts in Financial Networks} 

\author{Béni Egressy}
\author{Roger Wattenhofer}
\affil{ETH Zürich, Switzerland}
\affil{\textbf{\{begressy,wattenhofer\}@ethz.ch}}

\date{}

\maketitle

\begin{abstract}
We consider networks of banks with assets and liabilities. Some banks may be insolvent, and a central bank can decide which insolvent banks, if any, to bail out. We view bailouts as an optimization problem where the central bank has given resources at its disposal and an objective it wants to maximize. We show that under various assumptions and for various natural objectives this optimization problem is NP-hard, and in some cases even hard to approximate. Furthermore, we also show that given a fixed central bank bailout objective, banks in the network can make new debt contracts to increase their own market value in the event of a bailout (at the expense of the central bank). 
\end{abstract}

\newpage 
\section{Introduction}

With recent advances in decentralized finance, algorithmic trading and Fintech,\footnote{Financial technology (Fintech) is the technology and innovation that aims to compete with traditional financial methods in the delivery of financial services.} computer science is becoming more indispensable than ever to work and progress in finance. In this paper we look at central bank bailouts from a computer science perspective and present the first results that address the computational complexity of bailouts.

\begin{figure}[H]
\centering
\begin{tikzpicture}[scale=0.6, every node/.style={scale=1.0}, >={Triangle[width=2mm,length=1mm]},->]
    \node[circle, draw, minimum size=0.5cm] (S) at  (2,2.5) {$s$};
    \node[circle, draw, minimum size=0.5cm] (T) at  (5,0) {$t$};
    \node[circle, draw, minimum size=0.5cm] (U) at  (3,-3) {$u$};
    \node[circle, draw, minimum size=0.5cm] (D) at  (-1,0) {$d$};
    \node[circle, draw, minimum size=0.5cm] (V) at  (9,2) {$v$};
    \node[circle, draw, minimum size=0.5cm] (W) at  (9,-2) {$w$};
    \draw [semithick,->] (T) -- node[above] {\small $3$} (S);
    \draw [semithick,->] (S) -- node[above] {\small $4$} (D);
    \draw [semithick,->] (D) -- node[above] {\small $5$} (T);
    \draw [semithick,->] (D) -- node[above] {\small $5$} (U);
    \draw [semithick,->] (T) -- node[above] {\small $3$} (V);
    \draw [semithick,->] (T) -- node[above] {\small $3$} (W);
    \draw [semithick,->] (U) -- node[above] {\small $3$} (W);
    \node[rectangle, fill=white, draw, above right = -0.1cm of S] {\footnotesize $1$};
    \node[rectangle, fill=white, draw, below right = -0.1cm of T] {\footnotesize $4$};
    \node[rectangle, fill=white, draw, below right = -0.1cm of U] {\footnotesize $4$};
    \node[rectangle, fill=white, draw, below right = -0.1cm of D] {\footnotesize $5$};
    \node[rectangle, fill=white, draw, below right = -0.1cm of V] {\footnotesize $0$};
    \node[rectangle, fill=white, draw, below right = -0.1cm of W] {\footnotesize $0$};
\end{tikzpicture}
\caption{Example of a financial network. Nodes represent banks, edges represent liabilities (directed towards the creditor) and squares represent cash holdings. We can see that bank $d$ will default; even if it receives the full liability of $4$ from bank $s$, it will still be unable to pay off its debts ($5+5$) with the cash and incoming payments ($5+4$). This in turn means banks $u$ and $t$ will not be fully repaid. Bank $u$ will stay solvent, but bank $t$ will default, and this will also lead to bank $s$ defaulting. We see that in this network all the banks are dependent on bank $d$ to some extent. A well-directed bailout strategy could save bank $d$ at a cost of $1$, and as a result save all other defaulting banks in the network.}
\label{fig:intro_example_cascade}
\end{figure}
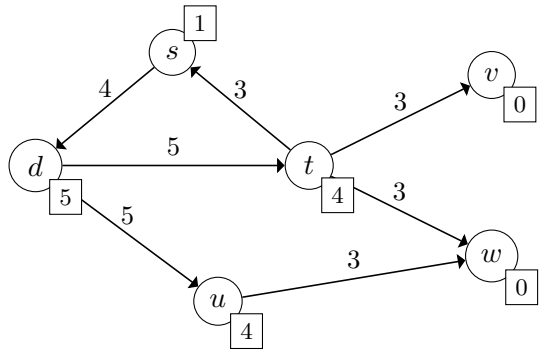

Financial institutions are linked in a complex network of contractual obligations, so when a shock\footnote{Examples of major shocks are the 2008 financial crisis and the 2020 COVID-19 pandemic.} hits, it can spread through the system. If an institution 
defaults on its 
debts, its creditors receive less payments than expected, and might also default.
In this way default can spread through the network, leading to chains of bankruptcies. This is often called a default cascade. The extent of such a cascade depends on the severity and location of the initial shock as well as the network topology. See Figure \ref{fig:intro_example_cascade} for a motivating example. 

In the wake of the 2008 financial crisis, many studies focused on understanding how the network topology affects default propagation and how resilient different network topologies are to various negative shocks. In this paper we want to look at what can be done to stop the cascade and mitigate the effects when a shock has already hit. 

During a financial crisis, one of the weapons in the arsenal of central banks is to carry out strategic bailouts.
By identifying the critical institutions to bail out, a central bank might be able to save the entire financial system at a low economic cost.  
In Appendix \ref{sec:too_big_to_fail} we challenge some conventional wisdom about bailouts. We show through examples why bailing out ``too-big-to-fail'' or ``too-central-to-fail'' banks might not be optimal. 
So how can a central bank identify the critical institutions it should bail out?
In this paper we formalize this question and frame it as an optimization problem. The goal is to understand this optimization problem from a computational perspective and analyze 
its complexity. 
We model the financial network as a directed graph, with nodes representing financial institutions and edges representing liabilities between them. Each node also has some external assets (cash). This network is the input for the central bank. The central bank has some objective that it wants to maximize. It can bail out banks in order to achieve this objective. We consider various natural choices for the central bank objective 
and present NP-hardness results. These results can be summarized in Table \ref{tab:NPhard_results}. They call into question the viability of making optimal bailout decisions in a complex financial network in the event of a crisis. 

In the final part of the paper we raise another issue with central bank bailouts. We look at the possibility of institutions using information about the central bank's objectives to their advantage. We show that given an imminent bailout scenario, banks can agree on new debt contracts that prove to be mutually beneficial when the central bank maximizes its objective. Moreover this is at the cost of the central bank. This means that a fixed, deterministic and publicly communicated bailout objective can be exploited by financial institutions.

\begin{table}[h]
\centering
\begin{tabular}{l l c l} 
 Objective & Budget & NP-hard\\ 
 \hline
 \hline
 max. total market value & unlimited & No & Theorem \ref{thm:central_unlimited} \\ 
 max. total market value & fixed & Yes & follows from Theorem \ref{thm:central_budget} \\
 \hline
 max. own market value & unlimited & Yes & Theorem \ref{thm:single_unlimited} \\ 
 max. own market value & fixed & Yes & follows from Theorem \ref{thm:single_unlimited} \\
 \hline
 max. no. banks saved & unlimited & No \\ 
 max. no. banks saved & fixed & Yes & Theorems \ref{thm:central_budget} and \ref{thm:central_budget_no_default_costs} \\
 \hline
 min. welfare loss* & unlimited & Yes & Theorem \ref{thm:welfare_loss} \\ 
 min. welfare loss & fixed & Yes & follows from Theorem \ref{thm:welfare_loss} \\
 \hline
 \\
\end{tabular}
\caption{Summary of our NP-hardness results. *Welfare loss balances maximizing the central bank's own value with minimizing the total default loss in the network. See Definition \ref{def:welfare_loss}.}
\label{tab:NPhard_results}
\end{table}
\vspace{-0.5cm}

\section{Literature Review}

Financial networks were introduced in the pioneering works of Allen and Gale \cite{AllenGale2000} and Eisenberg and Noe \cite{EisenbergNoe2001}. Many later works are either directly or indirectly based on the model introduced by Eisenberg and Noe. This model only assumes simple debt contracts between banks. The model has been extended to include default costs \cite{RogersVeraart2013}, cross-ownership, fire sales \cite{CifuentesFerrucciShin2005fireSales}, common assets \cite{elliott2014} and credit default swaps \cite{papp2020default,Schuldenzucker2016CDS}. 

Other recent developments on financial contagion were made by Gai and Kapadia \cite{Gai2010}, Glasserman and Young \cite{GlassermanYoung2015} and Battiston et al. \cite{Battiston2012debtrank}. For a survey on financial contagion see Glasserman and Young \cite{GlassermanYoung2016}. These works study how an initial shock spreads through a financial network based on the network topology and the inefficient liquidation of assets. In particular they explore what network structures lead to the greatest amplification of shocks and what regulations should therefore be put in place to avoid such structures forming. 

As a consequence, 
banks and regulatory bodies have acknowledged the importance of taking the full network into account when developing stress testing models, see Anand et al. \cite{Anand2014StressTestCanada}, Elsinger et al. \cite{elsinger2013Austria}, Gai et al. \cite{GaiKapadia2011}, Upper \cite{upper2011}. Nevertheless, 
Siebenbrunner and Sigmund \cite{siebenbrunner2018} find that financial contagion is still not priced into interbank markets.
Another line of work looks at what central banks can do to minimize the effects of contagion in a network. Freixas et al. \cite{FreixasParigiRochet2000} look at how liquidity shocks can be compensated by a central bank. 

Bernard, Capponi and Stiglitz \cite{BernardCapponiStiglitz2017} take a game theoretic perspective to analyze when banks would be incentivized to bail each other out rather than relying on the central bank.
They discuss bailouts as an optimization problem. However they 
focus on the game theoretic aspects. In their model, liabilities to the central bank are prioritised and the objective of the central bank is to minimize the welfare loss.
The authors are interested in whether banks are incentivized to bail each other out or whether, 
based on this central bank objective, 
they would rather wait for the central bank to bail out insolvent\footnote{We use ``insolvent'', ``bankrupt'' and ``in default'' interchangeably to mean an institution with more debts than assets. Legally speaking, bankruptcy is a process for liquidating what property and assets a debtor owns to pay off their debts, whereas insolvency is a financial state in which a person's (or company's) debts exceed their assets and default is when a debtor fails to make a payment in time.} banks. They assume a solution to the central bank's optimization problem can be found through 
exhaustive search. However, we show that this is unrealistic for large financial networks since the problem is NP-hard. 

The difficulty of finding optimal solutions to economic problems was noted already by Herbert Simon in his work on bounded rationality in 1982 \cite{herbert1982models}. He noted that human rationality is bounded, because there are limits to our thinking capacity, available information, and time. However to formalize such a statement one needs to turn to complexity theory. 
To this date there is limited research overlap in economics and computational complexity.
Arora et al. \cite{Arora2010} show that pricing derivatives can be computationally inefficient and Maymin \cite{maymin2011markets} argues that markets are efficient if and only if P$=$NP. In the context of financial networks, the recent works of Schuldenzucker et al. \cite{Schuldenzucker2016CDS,schuldenzucker2017} introduce a model incorporating CDSs, i.e., debt contracts where the payment obligation depends on the default of a specific other bank. They focus on the existence of a clearing solution in this model, and show that it is NP-hard to find a clearing vector. Papp et al. \cite{papp2020default} extend this result to show that approximating the ``best'' clearing vector within a $n$-factor is NP-hard. To our knowledge, the only result that discusses computational complexity in financial networks without CDSs is the work of Hemenway and Khanna \cite{hemenway2016} in the model of Elliot et al. \cite{elliott2014}, showing that among shocks of a given total size, it is NP-hard to find the one that is most damaging to the 
system in total.

Of course there is a vast branch of work in computer science on NP-hard problems and approximation algorithms. The results are too many to mention, but we will refer the reader to the results we rely on in this paper. 

\section{Preliminaries} 


Our model is based on the model of Rogers and Veraart \cite{RogersVeraart2013}. We consider a network with $n$ banks, $N:=\left\{u, v, w, ...\right\}$. Each bank may have liabilities to other banks in the network. We represent liabilities in the form of a matrix, $L \in \mathbb{R}^{n\times n}$, where the $uv$\textsuperscript{th} entry $l_{u,v}$ represents the liability of bank $u$ to bank $v$. We assume that $l_{u,v} \geq 0$ for all $u,v$ and $l_{u,u} = 0$ for all $u$. We denote the \emph{total liabilities} of bank $u$ by $L_{u} = \sum_{v \in N} l_{u,v}$. Each bank $u$ also has external assets (cash) of value $x_u \geq 0$. 
The network can be viewed as a directed graph, where banks are nodes and liabilities are directed edges. A tuple $(N, L, x)$ is called a \emph{financial network}.

Banks that cannot pay their total liabilities are said to be \emph{insolvent} or \emph{in default}. When a bank defaults, it first has to pay a share of its external assets and incoming payments to cover the costs associated with defaulting. In practise, this includes outstanding salaries, pension payments, clearing fees, losses due to inefficient asset liquidation, etc.
We account for this by introducing a default cost parameter $\beta \in [0,1]$. When a bank defaults, a $1-\beta$ share of its total assets is lost from the system, and the remaining $\beta$ share is distributed among its creditors.
Note that if $\beta = 1$, then we have no default costs. 

The \emph{recovery rate} $r_u$ of bank $u$ is the share of its liabilities it is able to pay. If bank $u$ is solvent, then $r_u = 1$, and if bank $u$ is insolvent then $r_u < 1$. The recovery rates of all the banks in the system can be represented as a vector $\vec{r} = (r_u)_{u \in N}$. The recovery rate is calculated after default costs 
so in fact if bank $u$ is insolvent, then $r_u < \beta$.

When a bank defaults, we assume that it pays out all of its external assets and incoming payments to its creditors. Moreover it makes the payments proportionally to the sizes of the liabilities: Payment of bank $u$ to bank $v$ is $p_{u,v}(\vec{r}) = r_u l_{u,v}$. Solvent banks pay all their liabilities in full. 
Given these principles, the recovery rate vector and payments satisfy: 
\begin{equation}
\label{equations}
    r_u = \min \left\{ \frac{a'_u}{L_u} , 1 \right\} \ , \ \ 
    p_{u,v} = r_{u} l_{u,v} \ , \ \ 
    a_u = x_u + \sum_{v \in N} p_{v,u} \ , \ \ 
    a'_u =
    \begin{cases}
        a_u & \text{if}\ a_u \geq L_u \\
        \beta a_u & \text{if}\ a_u < L_u
    \end{cases}
\end{equation}

In their paper \cite{EisenbergNoe2001}, Eisenberg and Noe show that under mild assumptions there is a unique clearing payment vector in a financial network with no default costs. Rogers and Veraart \cite{RogersVeraart2013} show that when default costs are added, the clearing vector might not be unique. They use a recursive algorithm to find the \emph{greatest clearing vector}. The greatest clearing vector is the clearing recovery rate vector such that no bank has a higher recovery rate in any clearing recovery rate vector.\footnote{Since an increase in the recovery rate of a bank cannot decrease the recovery rates of the other banks, the greatest clearing vector is well-defined and always exists. We refer to \cite{RogersVeraart2013} for details.} From now on we will assume we always take the greatest clearing vector as our solution if there is any ambiguity. This motivates the following definition. 

\begin{definition}[Clearing Recovery Rate Vector]
\label{def:clearing_vector}
Let $(N, L, x)$ be a financial network with default cost parameter $\beta$. Define the update function $f : [0,1]^n \rightarrow [0,1]^n$ as
\begin{align*}
    f_u(\vec{r})  : & = 
    \min \left\{ \frac{a'_u(\vec{r})}{L_u(\vec{r})} , 1 \right\} \\
    \vec{r}^{(k+1)} & = f(\vec{r}^{(k)})
\end{align*}
A recovery rate vector $\vec{r}$ is called \emph{clearing} 
if it is a fixed point of the update function $f$.
\end{definition}

Essentially the clearing recovery rate vector is defined as the fixed point of the banks iteratively transferring the assets they have (and receive) to their creditors until they have either paid their debts in full or 
have nothing left to give.
Clearing allows us to evaluate the value of each bank given the state of the network: It tells us what each bank would end up with if no more changes were made to the network and all existing liabilities were \emph{cleared}. 

\begin{definition}[Market Value]
\label{def:market_value}
We define the \emph{market value} of a bank $u$ as 
\begin{equation*}
    V_u = a'_u - r_u L_u = \max \left\{ a_u - L_u , 0 \right\}
\end{equation*}
\end{definition}
We denote the set of \emph{defaulting banks} by $N^{-} = \left\{ u \in N \mid a_u < L_u \right\}$. 

\begin{definition}[Shortfall, Exact Bailout]
\label{def:shortfall}
The \emph{shortfall} is the minimum amount a bank would need to receive to be saved from default. If a bank is solvent then this is $0$. We denote the shortfall of bank $u$ by
\begin{equation*}
    \Delta_u = \max \left\{ L_u - a_u , 0 \right\}
\end{equation*}
Note that as banks in the network are bailed out, bank $u$'s shortfall might decrease. A bailout of bank $u$ is considered \emph{exact} if it has market value exactly $0$ upon clearing after all bailouts.
\end{definition}

In addition to the $n$ banks in the system we also have a \emph{central bank}. The role of the central bank is to inject/distribute money in the system in order to optimize some objective. See Table \ref{tab:NPhard_results} for a list of objectives we consider. The central bank may be \emph{external} to the system or it may be one of the nodes in the financial network. Being external is equivalent to being a lone node in the network with no incoming or outgoing liabilities. Note that the money spent on bailouts has to be deducted to give the final value of the central bank. 

The clearing mechanism in Definition \ref{def:clearing_vector} allows us to compare two states of the network, e.g. one before bailouts and one after bailouts. 
The central bank essentially has the opportunity to inject money into the network to improve the outcome of the clearing process. 
For example if the objective is to save as many banks as possible, then the improvement would be the difference between ``the number of solvent banks upon clearing with the injection'' versus ``the number of solvent banks upon clearing without the injection''.



\begin{toappendix}
\section{Too Big to Fail?}
\label{sec:too_big_to_fail}

We start by challenging some conventional wisdom about bailout strategies. One expression that is often used is \emph{``too big to fail''}. Too big to fail describes a financial organisation or a business as so important to the economy that the central bank must take measures to prevent it from going bankrupt. This could be because many jobs directly depend on it, as in the case of the bailout of the auto industry in the US in 2009, or because many other businesses or financial institutions depend on it as in the case of the collapse of Bear Stearns in 2008. 

With Figure \ref{fig:indirect_bailout} we show an example of when bailing out the largest and most central bank directly might not be a good idea. Whatever the objective might be, 
it is cheaper to bail out bank $d$'s debtors, saving bank $d$ indirectly.
In Figure \ref{fig:too_big_to_fail} we show that it might not be the right strategy to bail out large and connected banks at all, neither directly nor indirectly. Bank $d$ is large in terms of assets as well as central to the network. However bailing out bank $d$ is very costly and primarily helps its biggest creditor, bank $s$, which already has a high market value.
Instead one can bail out the creditors of bank $d$ and let bank $d$ go bankrupt. 

\begin{figure}[H]
    \centering
    \begin{minipage}{0.48\textwidth}
        \centering
        \begin{tikzpicture}[scale=0.45, every node/.style={scale=1.0}, >={Triangle[width=2mm,length=1mm]},->]
            \node[circle, draw, minimum size=0.5cm] (A) at  (1,3) {$u$};
            \node[circle, draw, minimum size=0.5cm] (B) at  (0,0) {$v$};
            \node[circle, draw, minimum size=0.5cm] (C) at  (1,-3) {$w$};
            \node[circle, draw, minimum size=0.5cm] (D) at  (5,0) {$d$};
            \node[circle, draw, minimum size=0.5cm] (E) at  (9,2) {$s$};
            \node[circle, draw, minimum size=0.5cm] (F) at  (9,-2) {$t$};
            \draw [semithick,->] (A) -- node[above] {\small $6$} (D);
            \draw [semithick,->] (B) -- node[above] {\small $4$} (D);
            \draw [semithick,->] (C) -- node[above] {\small $10$} (D);
            \draw [semithick,->] (D) -- node[above] {\small $20$} (E);
            \draw [semithick,->] (D) -- node[below] {\small $30$} (F);
            \node[rectangle, fill=white, draw, below right = -0.1cm of A] (a) {\footnotesize $4$};
            \node[rectangle, fill=white, draw, below right = -0.1cm of B] (b) {\footnotesize $3$};
            \node[rectangle, fill=white, draw, below right = -0.1cm of C] (c) {\footnotesize $7$};
            \node[rectangle, fill=white, draw, below right = -0.1cm of D] (d) {\footnotesize $34$};
            \node[rectangle, fill=white, draw, below right = -0.1cm of E] (c) {\footnotesize $10$};
            \node[rectangle, fill=white, draw, below right = -0.1cm of F] (d) {\footnotesize $10$};
        \end{tikzpicture}
        \caption{Example of a financial network, where bailing out the largest bank (bank $d$) directly is not optimal. The default cost parameter is $\beta = 1/2$. The cost to bail out $d$ directly is $\Delta_d = (30+20)-(34 + \frac{1}{2}(4+3+7)) = 9$. But we can bail out banks $u$, $v$ and $w$ at a cost of $2+1+3 = 6$, which also saves bank $d$.
        In fact we can save bank $d$ at a minimum cost of $4$, by bailing out banks $v$ and $w$ only, such that $d$ receives $2+4+10 = 16$.}
        \label{fig:indirect_bailout}
    \end{minipage}\hfill
    \begin{minipage}{0.48\textwidth}
        \centering
        \begin{tikzpicture}[scale=0.45, every node/.style={scale=1.0}, >={Triangle[width=2mm,length=1mm]},->]
            \node[circle, draw, minimum size=0.5cm] (A) at  (0,2.5) {$u$};
            \node[circle, draw, minimum size=0.5cm] (B) at  (0,0) {$v$};
            \node[circle, draw, minimum size=0.5cm] (C) at  (0,-2.5) {$w$};
            \node[circle, draw, minimum size=0.5cm] (D) at  (5,0) {$d$};
            \node[circle, draw, minimum size=0.5cm] (E) at  (5,-4) {$s$};
            \node[circle, draw, minimum size=0.5cm] (F) at  (10,3.6) {$f$};
            \node[circle, draw, minimum size=0.5cm] (G) at  (10,1.8) {$g$};
            \node[circle, draw, minimum size=0.5cm] (H) at  (10,0) {$h$};
            \node[circle, draw, minimum size=0.5cm] (J) at  (10,-1.8) {$j$};
            \node[circle, draw, minimum size=0.5cm] (K) at  (10,-3.6) {$k$};
            \node[minimum size=0.5cm] (F1) at  (13.5,3.6) {};
            \node[minimum size=0.5cm] (G1) at  (13.5,1.8) {};
            \node[minimum size=0.5cm] (H1) at  (13.5,0) {};
            \node[minimum size=0.5cm] (J1) at  (13.5,-1.8) {};
            \node[minimum size=0.5cm] (K1) at  (13.5,-3.6) {};
            \draw [semithick,->] (A) -- node[above] {\small $30$} (D);
            \draw [semithick,->] (B) -- node[above] {\small $30$} (D);
            \draw [semithick,->] (C) -- node[above] {\small $30$} (D);
            \draw [semithick,->] (D) -- node[right] {\small $180$} (E);
            \draw [semithick,->] (D) -- node[above] {\small $4$} (F);
            \draw [semithick,->] (D) -- node[above] {\small $4$} (G);
            \draw [semithick,->] (D) -- node[above] {\small $4$} (H);
            \draw [semithick,->] (D) -- node[above] {\small $4$} (J);
            \draw [semithick,->] (D) -- node[above] {\small $4$} (K);
            \draw [semithick,->] (F) -- node[above right] {\small $12$} (F1);
            \draw [semithick,->] (G) -- node[above right] {\small $12$} (G1);
            \draw [semithick,->] (H) -- node[above right] {\small $12$} (H1);
            \draw [semithick,->] (J) -- node[above right] {\small $12$} (J1);
            \draw [semithick,->] (K) -- node[above right] {\small $12$} (K1);
            \node[rectangle, fill=white, draw, below right = -0.1cm of A] (a) {\footnotesize $50$};
            \node[rectangle, fill=white, draw, below right = -0.1cm of B] (b) {\footnotesize $50$};
            \node[rectangle, fill=white, draw, below right = -0.1cm of C] (c) {\footnotesize $50$};
            \node[rectangle, fill=white, draw, below right = -0.1cm of D] (d) {\footnotesize $10$};
            \node[rectangle, fill=white, draw, below right = -0.1cm of E] (c) {\footnotesize $500$};
            \node[rectangle, fill=white, draw, below right = -0.1cm of F] (f) {\footnotesize $10$};
            \node[rectangle, fill=white, draw, below right = -0.1cm of G] (g) {\footnotesize $10$};
            \node[rectangle, fill=white, draw, below right = -0.1cm of H] (h) {\footnotesize $10$};
            \node[rectangle, fill=white, draw, below right = -0.1cm of J] (j) {\footnotesize $10$};
            \node[rectangle, fill=white, draw, below right = -0.1cm of K] (k) {\footnotesize $10$};
        \end{tikzpicture}
        \caption{Example of a financial network, where it is better to not bail out the largest bank (bank $d$) at all. The default cost parameter is $\beta = 1/2$. The cost to bail out bank $d$ is $\Delta_d = 200 - 100 = 100$. On the other hand the cost to bail out banks $\{f, g, h, j, k\}$ is $5$ in total, since they each have total assets $10+1=11$ (incl. incoming assets) and total liabilities $12$.} 
        \label{fig:too_big_to_fail}
    \end{minipage}
\end{figure}


As a real world example, Bernard et al. \cite{BernardCapponiStiglitz2017} note that the ``bailout of AIG (during the financial crisis of 2008) was widely speculated to be an indirect bailout of Goldman Sachs, to which it had sold millions of dollars worth of insurance.''
If Goldman Sachs was close to bankruptcy, the situation might have been similar to Figure \ref{fig:indirect_bailout}, where an indirect bailout makes sense. On the other hand if the situation was closer to Figure \ref{fig:too_big_to_fail} and Goldman Sachs could have survived the losses, then this government bailout might have been suboptimal. 

Of course one might argue that such examples are convoluted, a network of liabilities 
would not look like this. Though this is most certainly the case, the examples indicate how important the network structure is, raising the important question of which banks and businesses to bail out in a crisis. How does one identify them? Could an algorithm do this?
\end{toappendix}

\section{Maximizing Market Value} \label{sec:unlimited_budget}

First we show that there is always an optimum solution where only insolvent banks receive money and insolvent banks are either bailed out exactly or not at all. As such we can reduce the problem to deciding which insolvent banks to bail out.

\begin{lemmarep}
\label{thm:reduction_to_binary_decision}
Consider a financial network $(N, L, x)$ with default cost parameter $\beta$. A (external) central bank with an unlimited budget wants to maximize the total market value
\begin{equation*}
    \sum_{u \in N} V_u - B
\end{equation*}
where $V_u$ is the market value of bank $u$ after clearing and $B$ is the total spend on bailouts.
There is an optimum solution where only insolvent banks receive money and they are either bailed out exactly
\footnote{See Definition \ref{def:shortfall}} 
or not at all.
\end{lemmarep}

\begin{proofsketch}
The idea of the proof is simple. 
First suppose there is an optimum solution where a bank receives more than its shortfall, then giving this bank exactly the shortfall amount will not change the objective value.
On the other hand suppose a bank receives less than its shortfall, but strictly more than $0$. Then instead we can take this amount, redistribute $\beta$ times this amount among the bank's creditors and keep the rest. This way the creditors end up with the same total assets, leading to the same clearing market values with a lower total bailout spend. This is a contradiction. 
See the appendix for details.
\end{proofsketch}

\begin{appendixproof}
Let $y \in [0,\infty)^n$ denote an allocation of bailout funds to the $n$ banks in the network. Then $B = \sum_u y_u$. Let $y^*$ be an optimum allocation with objective value $\sum_{u \in N} V^*_u - B^*$. We have to show that there is another optimum allocation $y'$ with $y'_u \in \{0, \Delta_u\}$ for all $u \in N$. The statement then follows.

Suppose there is some bank $v$, such that $y^*_v > \Delta_v$, i.e. bank $v$ receives more bailout money than its shortfall. Then let $y'_v = \Delta_v$. Since bank $v$ is still solvent when $y'_v = \Delta_v$, all of its creditors still receive their liabilities in full so all of the other banks in the network have the same market value, that is $V^*_u = V'_u$ for all $u \neq v$. And for $v$ we have $V'_v = V^*_v - y^*_v + \Delta_v$, but the total bailout spend, $B'$, is less by the same amount. Therefore we have $\sum_{u \in N} V'_u - B' =  \sum_{u \in N} V^*_u - B^*$

On the other hand suppose that there is some bank $v$, such that $0 < y^*_v < \Delta_v$. Then let $y'_v = 0$ and distribute $\beta y^*_v$ proportionally among $v$'s direct creditors. Since $y^*_v < \Delta_v$, bank $v$ is still insolvent in the optimum solution so it pays out $a_v + \beta y^*_v$ proportionally among its creditors. If instead we distribute $\beta y^*_v$ among the creditors, we ensure that each creditor ends up with the same total assets. Therefore the creditors and any other banks in the network satisfy the same clearing equations and must therefore have the same market value, i.e. $V'_u \geq V^*_u$ for all $u \neq v$. Moreover since bank $v$ is insolvent in either case, it has market value $V'_v = V^*_v = 0$. For $\beta<1$, this leads to a contradiction, since $\sum_{u \in N} V'_u  =  \sum_{u \in N} V^*_u$ and $B' < B^*$, but we assumed $y*$ is an optimum solution. And for $\beta = 1$ we get an optimum solution with $y'_v = 0$. 

Therefore there is an optimum allocation $y'$ with $y'_u \in \{0, \Delta_u\}$ for all $u \in N$.
\end{appendixproof}

One can easily adapt the proof to the other objectives we consider. The essence of the proof is the same; a solvent bank does not need more money and instead of a bank receiving a partial bailout it is better to distribute this amount among the bank's creditors. 

Now we show that if a central bank with an unlimited budget wants to maximize the total value (minus bailout costs) in the network, then finding an optimum strategy is trivial. 

\begin{theorem1rep}
\label{thm:central_unlimited}
Consider a financial network $(N, L, x)$ with default cost parameter $\beta$. A (external) central bank with an unlimited budget wants to maximize 
\begin{equation*}
    \sum_{u \in N} V_u - B
\end{equation*}
where $V_u$ is the market value of bank $u$ after clearing and $B$ is the total spend on bailouts.
Bailing out all insolvent banks is optimal. 
Moreover for $\beta<1$, $B$ minimal such that the objective is maximized is given by $B = \sum_{u \in N} \Delta_u$, where the shortfalls are calculated assuming all incoming liabilities are paid in full.
\end{theorem1rep}

\begin{proofsketch}
Note that since value can only escape from the network through default costs, the total market value in the system is given by the total value of the external assets minus the total default costs. If the central bank bails out all the insolvent banks then default losses will be zero. Furthermore any money injected into the system will add to the total market value, offsetting exactly the additional bailout cost.
See the appendix for details.
\end{proofsketch}

\begin{appendixproof}
Let $N^{-}$ denote the set of insolvent banks. Note that since value can only escape from the network through default costs, the total market value in the system is given by the total value of the external assets minus the total default costs. 
\begin{equation*}
    \sum_{u \in N} V_u = \sum_{u \in N} x_u - \left( 1 - \beta \right) \sum_{u \in N^{-}} a_u
\end{equation*}
If the central bank bails out all the insolvent banks then default losses will be zero. Furthermore any money injected into the system will add to the total market value directly, since value can no longer escape from the system. As in Definition \ref{def:shortfall} the bailout amount for bank $u$ is $\Delta_u$. Therefore after central bank bailouts we have  
\begin{align*}
    \sum_{u \in N} V_u - B 
    & = \sum_{u \in N} (x_u + \Delta_u) - \sum_{u \in N} \Delta_u \\
    & = \sum_{u \in N} x_u 
\end{align*}
One can see that injecting any additional money into the system will cancel out exactly. It adds to the total value, but is subtracted again through $B$. Moreover, for $\beta<1$, one can see that if any bank is left to default, then some value will be lost from the system and $\sum_{u \in N} V_u - B$ will be lower. For $\beta = 1$, any allocation is an optimal solution, since no value leaves the system. Therefore bailing out all insolvent banks is always an optimal solution. 

To see why $B$ minimal is given by $B = \sum_{u \in N} \Delta_u$, 
first note that upon bailing out a bank, the shortfalls of other insolvent banks in the network may decrease. 
Given that all banks will be bailed out in an optimum solution, we can calculate the banks' assets based on them receiving their incoming liabilities in full. This is equivalent to calculating with $\beta = 1$. If $B$ were any lower, then some bank would default and value would be lost through default costs. 

Note that for $\beta=1$, i.e. no default costs, $B=0$ is minimal such that the objective is maximized, with no bank receiving any money.
\end{appendixproof}

Now consider the central bank as a node in the financial network, with external assets and liabilities. The central bank now wants to maximize its own market value. Alternatively, consider any bank in the system that wants to maximize its own market value through strategic bailouts, sometimes referred to as \emph{bail-ins}.
Note that the total amount the bank spends on the bailouts $B$, is deducted from its assets (and therefore market value).
We show that this problem is NP-hard. 

\begin{theorem1}
\label{thm:single_unlimited}
Consider a financial network $(N, L, x)$. A bank $u$ wants to maximize its own market value through bailouts. That is, bank $u$ wants to maximize
\begin{equation*}
    V_u = \max \left\{ x_u + \sum_{v \in N} r_{v} l_{v,u} - L_u - B, 0 \right\}.
\end{equation*}
For any default cost parameter $\beta<1$, deciding which banks to bail out is NP-hard.
\end{theorem1}

Note that for $\beta=1$ the problem is not NP-hard. When there are no default costs, there is no gain 
from bailing out insolvent institutions; no default costs are saved from leaving the system and the bailout money is simply distributed proportionally among the creditors. The best a bank that bails out another bank can hope for is that the exact bailout amount flows back to it upon clearing. Thus one optimum solution is to not bail out any banks at all.

\begin{proof}
To show that the optimization problem is NP-hard, we prove that the corresponding decision problem is NP-complete. The corresponding decision problem is:
\begin{itemize}
    \item[] Instance: Tuple $(N, L, x)$, an index indicating the (central) bank $0$, and a real number $k'$
    \item[] Question: Can bank $0$ achieve a market value of at least $k'$ through (strategic) bailouts?
\end{itemize}
We show that for any default cost parameter $\beta<1$, this decision problem is NP-complete by a reduction from \textsc{Vertex Cover}, a well known NP-complete problem \cite{Karp1972}. Let $(G=(V,E), k)$ be an instance of the vertex cover decision problem. 
The question is whether a vertex cover of size at most $k$ exists. 
Given an instance of vertex cover, $(G=(V,E), k)$, we construct an instance of the bailout decision problem as follows: 
\begin{itemize}
    \item Let $n = \abs*{V}$ and let $d_v$ be the degree of $v$ in $G$.
    \item We add bank $0$ with external assets $x_0 = n$.
    \item We define $\epsilon = \frac{1}{2}\left( 1 - \beta \right)$. For each $e \in E$ we add an \emph{edge bank} $e$ with external assets $x_e = 0$ and debt $l_{e,0} = 2\beta + 2\epsilon = 1 + \beta$.
    \item For each $v \in V$ we add a \emph{vertex bank} $v$ with external assets $x_{v} = d_v$ and debts $l_{v,e} = 1 + \frac{1}{d_v}\epsilon$ for all edges $e$ that are incident to $v$.
    \item $k' = n + (1+\beta)\abs*{E} - k\epsilon$
\end{itemize}

See Figure \ref{fig:bank_vertex_cover} for an illustration. 
Note that initially every bank in this financial network (apart from bank $0$) is insolvent. Each vertex bank $v$ has total assets $a_v = d_v$ and total liabilities $L_v = d_v \left( 1 + \frac{1}{d_v}\epsilon \right) = d_v + \epsilon$.
So each bank $v$ can be bailed out at a cost of $L_v - a_v = \epsilon$.
Since each bank $v$ is insolvent, without a bailout it will pay out $a'_v = \beta a_v = \beta d_v$ after deduction of default costs, so each incident edge bank will receive exactly $\beta$.
Therefore every edge bank $e$ initially receives $\beta$ from each of its incident vertices giving total assets $a_e = 2\beta$ and total debt $L_e = 2\beta+2\epsilon$, resulting in a shortfall of $2\epsilon$.

\begin{figure}[H]
\centering
\begin{tikzpicture}[scale=0.6, every node/.style={scale=1.0}, >={Triangle[width=2mm,length=1mm]},->]
    \node[circle, draw, minimum size=0.5cm] (0) at  (14,1) {$0$};
    \node[circle, draw, minimum size=0.5cm] (e1) at  (7,4) {$e_1$};
    \node[circle, draw, minimum size=0.5cm] (e2) at  (7,2) {$e_2$};
    \node[circle, draw, minimum size=0.5cm] (e3) at  (7,0) {$e_3$};
    \node (e4) at  (7,-2.4) {...};
    \node[circle, draw, minimum size=0.5cm] (v1) at  (0,4) {$v_1$};
    \node[circle, draw, minimum size=0.5cm] (v2) at  (0,0.4) {$v_2$};
    \node (v3) at  (0,-2.4) {...};
    \draw [semithick,->] (e1) -- node[above] {\small $1+\beta$} (0);
    \draw [semithick,->] (e2) -- node[above] {\small $1+\beta$} (0);
    \draw [semithick,->] (e3) -- node[above] {\small $1+\beta$} (0);
    \draw [semithick,->] (e4) -- (0);
    \draw [semithick,->] (v1) -- node[above] {\small $1 + \frac{\epsilon}{d_{v_1}}$} (e1);
    \draw [semithick,->] (v1) -- node[above] {\small $1 + \frac{\epsilon}{d_{v_1}}$} (e2);
    \draw [semithick,->] (v1) -- node[above] {\small $1 + \frac{\epsilon}{d_{v_1}}$} (e3);
    \draw [semithick,->] (v2) -- node[below] {\small $1 + \frac{\epsilon}{d_{v_2}}$} (e2);
    \draw [semithick,->] (v2) -- node[above] {\small $1 + \frac{\epsilon}{d_{v_2}}$} (e4);
    \draw [semithick,->] (v3) -- (e4);
    \node[rectangle, fill=white, draw, below right = -0.1cm of 0] (d) {\footnotesize $n$};
    \node[rectangle, fill=white, draw, below right = -0.1cm of e1] (a) {\footnotesize $0$};
    \node[rectangle, fill=white, draw, below right = -0.1cm of e2] (a) {\footnotesize $0$};
    \node[rectangle, fill=white, draw, below right = -0.1cm of e3] (a) {\footnotesize $0$};
    \node[rectangle, fill=white, draw, below right = -0.1cm of v1] (b) {\footnotesize $d_{v_1}$};
    \node[rectangle, fill=white, draw, below right = -0.1cm of v2] (b) {\footnotesize $d_{v_2}$};
\end{tikzpicture}
\caption{The construction used in Theorem \ref{thm:single_unlimited} to give a reduction from \textsc{Vertex Cover} to the single bank market value optimization problem.} 
\label{fig:bank_vertex_cover}
\end{figure}

For the easy direction of the proof, let $S$ be a vertex cover in $G$ of size at most $k$. Then in the financial network, bank $0$ bails out all the vertex banks corresponding to the vertices in $S$. In this way, all the edge banks will have at least one solvent debtor and so will have incoming assets greater than $1+\beta$ (more than $1$ from the solvent vertex bank and at least $\beta$ from the other). Thus each edge bank will be solvent and bank $0$ will have market value $V_0 = n + (1+\beta)\abs*{E} - \epsilon \abs*{S} \geq n + (1+\beta)\abs*{E} - k\epsilon = k'$.

On the other hand, suppose there is a set of banks $T$, that bank $0$ can bail out to reach a market value of at least $k' = (1+\beta)\abs*{E} - k\epsilon$. We claim that given this set $T$, we can find a set $T'$ with no edge banks, such that if bank $0$ bails out the banks in $T'$, then all edge banks are solvent and bank $0$ reaches a market value of at least $k'$. 

Since each bank $e$ is initially insolvent, bank $0$ receives $2 \beta^2 < 2\beta$ from each edge bank. But if bank $0$ bails out a bank $e$ for $2\epsilon$ then it receives $2\beta + 2\epsilon$, thereby increasing its market value. 
Therefore if some edge bank $e$ is not solvent when bailing out the banks in $T$, then bank $0$ could increase its market value by bailing out $T \cup \{e\}$ instead.

However note that it is always better to bail out one of the incident vertex banks. Let $e=(u,v)$. Bank $0$ can bail out bank $v$ for $\epsilon$ and then $e$ will now receive more than $1$ from $v$ and $\beta$ from $u$ so it 
stays solvent and pays off the whole debt to bank $0$. It costs only $\epsilon$ to bail out $v$, instead of the $2\epsilon$ that it costs to bail out $e$ directly.

So to get $T'$ from $T$, we can first add any insolvent edge banks, and then for each edge bank in the set, we can replace it with one of its incident vertex banks (in case neither is in the set so far). So we have a set $T'$ of vertex banks and the vertices corresponding to these banks form a vertex cover in $G$. Moreover since the market value $V_0 = n + (1+\beta)\abs*{E} - \epsilon \abs*{T'} \geq k'$, we must have $\abs*{T'} \leq k$, so the size of the vertex cover must be at most $k$.
\end{proof}
The proof can be adapted to many other objective functions as well as other variants of the model. Below we consider saving the highest number of banks with a fixed bailout budget.

\section{Bailouts on a Budget}

\begin{theorem1rep}
\label{thm:central_budget}
Consider a financial network $(N, L, x)$. A central bank with a fixed budget $B$ wants to maximize the number of insolvent banks it saves. For any default cost parameter $\beta<1$, this optimization problem is NP-hard. Moreover assuming the exponential time hypothesis (ETH), there is no polynomial time algorithm that approximates this problem within a factor of $n^{1/(log log n)^c}$ for some universal constant $c>0$ independent of $n = \abs*{N}$.
\end{theorem1rep}

There are many natural reductions that we could choose here. We choose the \textsc{Densest} $k$-\textsc{Subgraph} problem as it gives the strongest inapproximability result. Choosing the best set of vertex banks to bail out (in order to save the most edge banks) will correspond to choosing the set of vertices in $G$ that induces the subgraph with the most edges. Note that the inapproximability factor, $n^{1/(log log n)^c}$, is super-constant. So the problem cannot be approximated within a constant factor for any constant.

\begin{proofsketch}
We prove this by reduction from the \textsc{Densest} $k$-\textsc{Subgraph} problem. In the densest $k$-subgraph problem, we are given an undirected graph $G$ on $n$ vertices and a positive integer $k \leq n$. The goal is to find a set $S$ of $k$ vertices such that the induced subgraph on $S$ has the most edges.
The construction we use is similar to the one used in Theorem \ref{thm:single_unlimited}. However we now choose the liabilities such that both banks $u$ and $v$ have to be bailed out in order to save bank $e$, where $e=(u,v)$ in the original graph. 
See the appendix for details.
\end{proofsketch}

\begin{appendixproof}
We prove this by reduction from the \textsc{Densest} $k$-\textsc{Subgraph} problem. 
In the densest $k$-subgraph problem, we are given an undirected graph $G$ on $n$ vertices and a positive integer $k \leq n$. The goal is to find a set $S$ of $k$ vertices such that the induced subgraph on $S$ has the highest number of edges. The decision problem asks, whether there is a subset of $k$ vertices such that the induced subgraph has at least $m$ edges.

The bailout decision problem asks whether at least $m'$ banks can be saved with a bailout budget of $B$. 
The construction we use for the reduction is similar to the one used in Theorem \ref{thm:single_unlimited}. However we now choose the liabilities such that both banks $u$ and $v$ have to be bailed out in order to save bank $e$, where $e=(u,v)$ in the original graph. We construct the financial network as follows:
\begin{itemize}
    \item We add bank $0$ with external assets $x_0 = 0$.
    \item For each $e \in E$ we add a bank $e$ with external assets $x_e = 0$ and debt $l_{e,0} = 2$.
    \item We define $\epsilon = \frac{1}{3}\left( 1 - \beta \right)$. For each $v \in V$ we add a node $v$ with external assets $x_{v} = d_v$ and debts $l_{v,e} = 1 + \frac{1}{d_v}\epsilon$ for all edges $e$ that are incident to $v$.
    \item We set the budget at $B = k\epsilon$, and $m'=k+m$
\end{itemize}
See Figure \ref{fig:bank_densest_subgraph_1} for an illustration.

\begin{figure}
\centering
\begin{tikzpicture}[scale=0.6, every node/.style={scale=1.0}, >={Triangle[width=2mm,length=1mm]},->]
    \node[circle, draw, minimum size=0.5cm] (0) at  (14,1) {$0$};
    \node[circle, draw, minimum size=0.5cm] (e1) at  (7,4) {$e_1$};
    \node[circle, draw, minimum size=0.5cm] (e2) at  (7,2) {$e_2$};
    \node[circle, draw, minimum size=0.5cm] (e3) at  (7,0) {$e_3$};
    \node (e4) at  (7,-3) {...};
    \node[circle, draw, minimum size=0.5cm] (v1) at  (0,4) {$v_1$};
    \node[circle, draw, minimum size=0.5cm] (v2) at  (0,0) {$v_2$};
    \node (v3) at  (0,-3) {...};
    \draw [semithick,->] (e1) -- node[above] {\small $2$} (0);
    \draw [semithick,->] (e2) -- node[above] {\small $2$} (0);
    \draw [semithick,->] (e3) -- node[above] {\small $2$} (0);
    \draw [semithick,->] (e4) -- (0);
    \draw [semithick,->] (v1) -- node[above] {\small $1 + \frac{\epsilon}{d_{v_1}}$} (e1);
    \draw [semithick,->] (v1) -- node[above] {\small $1 + \frac{\epsilon}{d_{v_1}}$} (e2);
    \draw [semithick,->] (v1) -- node[above] {\small $1 + \frac{\epsilon}{d_{v_1}}$} (e3);
    \draw [semithick,->] (v2) -- node[below] {\small $1 + \frac{\epsilon}{d_{v_2}}$} (e2);
    \draw [semithick,->] (v2) -- node[above] {\small $1 + \frac{\epsilon}{d_{v_2}}$} (e4);
    \draw [semithick,->] (v3) -- (e4);
    \node[rectangle, fill=white, draw, below right = -0.1cm of 0] (d) {\footnotesize $0$};
    \node[rectangle, fill=white, draw, below right = -0.1cm of e1] (a) {\footnotesize $0$};
    \node[rectangle, fill=white, draw, below right = -0.1cm of e2] (a) {\footnotesize $0$};
    \node[rectangle, fill=white, draw, below right = -0.1cm of e3] (a) {\footnotesize $0$};
    \node[rectangle, fill=white, draw, below right = -0.1cm of v1] (b) {\footnotesize $d_{v_1}$};
    \node[rectangle, fill=white, draw, below right = -0.1cm of v2] (b) {\footnotesize $d_{v_2}$};
\end{tikzpicture}
\caption{Figure shows the construction used in Theorem \ref{thm:central_budget} to give a reduction from the \textsc{Densest} $k$-\textsc{Subgraph} problem to the problem of saving the most banks on a limited budget.} 
\label{fig:bank_densest_subgraph_1}
\end{figure}

Note that initially every bank (apart from bank $0$) in this financial network is insolvent. Note also that each bank $v$ has the same assets and liabilities as in Theorem \ref{thm:single_unlimited}. So in particular it has shortfall $\epsilon$ and initially pays out exactly $\beta$ to each incident edge bank.

An bank $e$ initially receives $\beta$ from each of its debtors giving total assets of $a_e = 2\beta$ and total debt of $L_e = 2$, resulting in a shortfall of $2 - 2\beta = 6\epsilon$. Notice that it is cheaper to bail out both corresponding vertex banks than to bail out bank $e$ directly. Moreover if one of the vertex banks has been bailed out, bank $e$ is still insolvent and its shortfall is at least 
\begin{equation*}
    2 - (1 + \epsilon + \beta) = 1 - \beta - \epsilon = \frac{2}{3} \left( 1 - \beta \right) > \epsilon 
\end{equation*}
So also when one of the vertex banks has already been bailed out it is better to bail out the other rather than to bail out bank $e$ directly. And if $u$ and $v$ are both bailed out, then bank $e$ is already solvent. Therefore in an optimal solution only vertex banks are bailed out.

Now suppose there is a solution, $S$, to the densest $k$-subgraph decision problem. Then we can bail out the banks corresponding to these vertices at a total cost of $k\epsilon \leq B$, saving a total of at least $k+m=m'$ banks ($k$ vertex and at least $m$ edge banks). 

On the other hand suppose there is a solution, $T$, for the bailout decision problem. Then, as in the proof of Theorem \ref{thm:single_unlimited}, we show that there is another solution, $T'$, with at most $k$ vertex banks, such that at least $m$ edge banks are also saved. We have argued above, that it is better to bailout both incident vertex banks than to bailout an edge bank directly, so we can replace any edge bank in $T$ with vertex banks to give $T'$. Then the cost of bailing out $T'$ is less than the cost of bailing out $T$, which was less than $B=k\epsilon$. Since the cost of bailing out a vertex bank is $\epsilon$, we can also conclude that $\abs*{T'} \leq k$, and at least $m$ vertex banks must also be saved. Therefore the vertices in $G$ corresponding to $T'$ must be a solution to the densest k-subgraph problem.


This proves that the problem of saving the most banks on a fixed budget is an NP-hard problem. But we can say more than this using the inapproximability result for \textsc{Densest} $k$-\textsc{Subgraph} \cite{densestksubgraph2017}. Assuming the exponential time hypothesis (ETH), there is no polynomial time algorithm that approximates densest-k subgraph to within a $n^{1/(log log n)^c}$ factor of the optimum. 

In our construction, the shortfall of each vertex bank is $\epsilon$ so with the given budget of $k\epsilon$ the central bank can bail out exactly $k$ vertex banks. The optimal solution is therefore to bail out the $k$ vertex banks that 2-cover the most edge banks. By 2-covering edge bank $e$ for $e=(u,v)$ we mean that both $u$ and $v$ are bailed out. This corresponds to finding the densest k-subgraph in the original graph $G$.  
Note that the total banks that are bailed out includes the $k$ vertex banks, and the inapproximability for \textsc{Densest} $k$-\textsc{Subgraph} only carries over to the number of edge banks. However, the number of edge banks is also at least $\Omega (k)$,\footnote{This may not hold for very sparse graphs, but these cases are easy to solve exactly.} so the inapproximability carries over to the total number of banks bailed out. So assuming ETH there can also be no polynomial time algorithm that approximates our problem to within a $n^{1/(log log n)^c}$ factor of the optimum. 
\end{appendixproof}

This result can be extended to many other objectives.
See Theorem \ref{thm:theorem_fixed_max_value} in the Appendix for the case of maximizing total market value on a budget. The NP-hardness result can also be extended to financial networks that include hierarchical debt (generalisation of senior and junior debts to more seniority levels) and debt contracts with due dates. However the inapproximability result might not translate to different cases and objectives. 

\begin{toappendix}
\begin{theorem}
\label{thm:theorem_fixed_max_value}
Consider a financial network $(N, L, x)$. A central bank with a fixed budget $B$ wants to maximize the total market value,
\begin{equation*}
    \sum_{u \in N} V_u
\end{equation*}
where $V_u$ is the market value of bank $u$ after clearing and $B$ is the total spend on bailouts. For any default cost parameter $\beta<1$, this optimization problem is NP-hard.
\end{theorem}

\begin{proofsketch}
This proof follows from the proof of Theorem \ref{thm:central_budget} with a minor modification to the construction. Instead of no external assets, each edge bank now has external assets $x_e = 2n$ and debt $l_{e,0} = 2n + 2$. This way we ensure that saving an edge bank is more valuable than saving any two vertex banks (without indirectly also saving an edge bank). On the other hand it still holds, that it is cheaper to save the corresponding two vertex banks rather than to save an edge bank directly. This ensures that an optimal solution still bails out only vertex banks and tries to ``cover'' as many edge banks with these vertex banks as possible. That is, we still have a reduction from the \textsc{Densest} $k$-\textsc{Subgraph} problem.
\end{proofsketch}
\end{toappendix}

Theorem \ref{thm:central_budget} leaves an open question: Is the problem NP-hard for $\beta=1$, i.e., no default costs? The same construction does not work, since all the banks would be solvent, and even with different liabilities we cannot force the optimum solution to only bail out vertex banks directly. 
However we show in the following that the problem is indeed still NP-hard.

\begin{theorem1rep}
\label{thm:central_budget_no_default_costs}
Consider a financial network $(N, L, x)$ with no default costs, i.e., $\beta = 1$. A central bank with a fixed budget $B$ wants to maximize the number of insolvent banks it bails out. This optimization problem is NP-hard.
\end{theorem1rep}

\begin{proofsketch}
We prove this by reduction from the \textsc{Maximum Independent Set} problem. 
In the maximum independent set problem, we are given an undirected graph $G$ on $n$ vertices. An \emph{independent set} is a set $S$ of vertices such that no two vertices in $S$ are adjacent. The goal is to find an independent set of largest possible size.
The construction we use is similar to the one used in Theorem \ref{thm:single_unlimited}, but the argument is quite different. Rather than trying to use as few vertices as possible to cover the edges, this time we want to choose as many vertices as possible without common edges. We first argue that an optimum solution will save all edge banks, and then reallocate the bailout money from the edge banks to as many vertex banks as possible (without overlap). This money will reach (and save) the edge banks anyway so we can freely reallocate it. See the appendix for details.
\end{proofsketch}

\begin{appendixproof}
We prove this by reduction from the \textsc{Independent Set} decision problem. 
In the \textsc{Independent Set} decision problem, we are given an undirected graph $G$ on $n$ vertices. An \emph{independent set} is a set $S$ of vertices such that no two vertices in $S$ are adjacent. The goal is to decide whether an independent set of size $k$ exits.

The construction we use for the reduction is similar to the one used in Theorem \ref{thm:single_unlimited}.
We construct the following financial network:
\begin{itemize}
    \item We add bank $0$ with external assets $x_0 = 0$.
    \item We add bank $1$ with external assets $x_1 = 0$.
    \item For each $e \in E$ we add a bank $e$ with external assets $x_e = 0$ and debt $l_{e,0} = 1$.
    \item For each $v \in V$ we add a node $v$ with external assets $x_{v} = 0$, debts $l_{v,e} = 1$ for all edges $e$ that are incident to $v$, and debt $l_{v,1} = 1$.
\end{itemize}

The central bank receives a budget of $\abs{E}+k$ and wants to maximize the number of banks that it saves from bankruptcy. See Figure \ref{fig:bank_densest_subgraph} for an illustration. We denote the set of insolvent banks by $N^{-}$, the ``edge'' banks by $E'$ and the ``vertex'' banks by $V'$, such that $N^{-} = E' \cup V'$. We show that deciding whether the central bank can save $\abs{E}+k$ banks is as hard as deciding whether there is an independent set of size $k$ in $G$.

\begin{figure}
\centering
\begin{tikzpicture}[scale=0.6, every node/.style={scale=1.0}, >={Triangle[width=2mm,length=1mm]},->]
    \node[circle, draw, minimum size=0.5cm] (0) at  (14,1) {$0$};
    \node[circle, draw, minimum size=0.5cm] (1) at  (-6,1) {$1$};
    \node[circle, draw, minimum size=0.5cm] (e1) at  (7,4) {$e_1$};
    \node[circle, draw, minimum size=0.5cm] (e2) at  (7,2) {$e_2$};
    \node[circle, draw, minimum size=0.5cm] (e3) at  (7,0) {$e_3$};
    \node (e4) at  (7,-3) {...};
    \node[circle, draw, minimum size=0.5cm] (v1) at  (0,4) {$v_1$};
    \node[circle, draw, minimum size=0.5cm] (v2) at  (0,0) {$v_2$};
    \node (v3) at  (0,-3) {...};
    \draw [semithick,->] (e1) -- node[above] {\small $1$} (0);
    \draw [semithick,->] (e2) -- node[above] {\small $1$} (0);
    \draw [semithick,->] (e3) -- node[above] {\small $1$} (0);
    \draw [semithick,->] (e4) -- (0);
    \draw [semithick,->] (v1) -- node[above] {\small $1$} (e1);
    \draw [semithick,->] (v1) -- node[above] {\small $1$} (e2);
    \draw [semithick,->] (v1) -- node[above] {\small $1$} (e3);
    \draw [semithick,->] (v2) -- node[below] {\small $1$} (e2);
    \draw [semithick,->] (v2) -- node[above] {\small $1$} (e4);
    \draw [semithick,->] (v3) -- (e4);
    \draw [semithick,->] (v1) -- node[above] {\small $1$} (1);
    \draw [semithick,->] (v2) -- node[above] {\small $1$} (1);
    \draw [semithick,->] (v3) -- (1);
    \node[rectangle, fill=white, draw, below right = -0.1cm of 0] (d) {\footnotesize $0$};
    \node[rectangle, fill=white, draw, below right = -0.1cm of e1] (a) {\footnotesize $0$};
    \node[rectangle, fill=white, draw, below right = -0.1cm of e2] (a) {\footnotesize $0$};
    \node[rectangle, fill=white, draw, below right = -0.1cm of e3] (a) {\footnotesize $0$};
    \node[rectangle, fill=white, draw, below right = -0.1cm of v1] (b) {\footnotesize $0$};
    \node[rectangle, fill=white, draw, below right = -0.1cm of v2] (b) {\footnotesize $0$};
\end{tikzpicture}
\caption{Figure shows the construction used in Theorem \ref{thm:central_budget} to give a reduction from the \textsc{Densest} $k$-\textsc{Subgraph} problem to the problem of saving the most banks on a limited budget.} 
\label{fig:bank_densest_subgraph}
\end{figure}

Note that initially every bank (apart from banks $0$ and $1$) in this financial network is insolvent. Each vertex bank $v$ has zero assets and liabilities $d_v + 1$, where $d_v$ is the degree of $v$ in $G$. Saving a bank $v$ automatically saves all of its incident edge banks, since $v$ now pays off its debt of $1$ to each one of them and they each have total liabilities of $1$. Therefore at a cost of $d_v + 1$, the central bank can save $d_v + 1$ institutions. This is a cost of $1$ per bank. On the other hand saving edge banks directly also costs $1$ per bank saved. Thus it is easy to see that it is not possible to save any number of banks at an average cost below $1$. 


Now consider saving vertex banks $u$ and $v$, where $e=(u,v)$ is an edge in graph $G$. Then the total cost for saving $u$ and $v$ is $d_u + d_v + 2$ and $d_v + d_u + 1$ banks are saved (since they share the edge bank $e=(u,v)$). This cost is above $1$ per bank saved, and since it is not possible to save banks below an average cost of $1$, we can conclude that $u$ and $v$ cannot both be in the set of saved banks. 

We conclude that $\abs{E}+k$ banks can be saved at cost $\abs{E}+k$ if and only if there is an independent set of size $k$ in $G$. Let $S' \subseteq N^{-}$ be a set of $\abs{E}+k$ banks saved at a cost of $\abs{E}+k$, then the vertices corresponding to $S' \cap V'$ form an independent set of size at least $k$ in $G$. To get an independent set of size exactly $k$ we can take any subset of these vertices of size $k$. On the other hand let $S$ be an independent set of size $k$ in $G$, then the central bank can save the $k$ banks corresponding to $S$ in addition to all the remaining insolvent edge banks, thereby saving $\abs{E}+k$ banks at a total cost of $\abs{E}+k$. 
\end{appendixproof}

It is worth noting that real-world instances might be small enough to be tractable. Certainly if only a few major banks are considered then this will be the case, but as we start adding more financial institutions, companies and private individuals, the network grows and the runtime grows super-polynomially with it.

\section{Minimizing Welfare Loss}

Inspired by other models in the literature \cite{BernardCapponiStiglitz2017}, we extend our basic model of a financial network with senior and junior debts. Furthermore we add the central bank as a node to the network and consider the objective of minimizing the \emph{welfare loss}. 

\begin{definition}[Senior Debt]
    A bank may have \emph{senior} and \emph{junior} creditors. Payments to the senior creditor are prioritised. That is, if a bank is in default, then it uses its assets to first pay off debts (proportionally) among its senior creditors before sharing any remaining assets proportionally among its junior creditors. 
\end{definition}

In our model only the central bank will be considered a senior creditor. This is in line with \cite{BernardCapponiStiglitz2017}. More importantly this is in line with the federal priority statute.\footnote{The federal priority statute, 31 U.S.C. § 3713,[FN1] provides that, when a debtor of the United States is insolvent and not in bankruptcy, it must pay its debts to the government before any other creditor.} When considering senior debt, the clearing vector has to be calculated accordingly. 
See Appendix \ref{app:clearing_w_seniors} for details on how equations (\ref{equations}) are adapted. If bank $u$ defaults, we denote by $\delta_u$ the loss that the central bank endures from $u$'s default,
\begin{equation*}
    \delta_u = l_{u,0} - p_{u,0} = \max \left\{ l_{u,0} - \beta a_u, 0 \right\}.
\end{equation*}
Now we can define the welfare loss.
\begin{toappendix}
\label{app:clearing_w_seniors}
We extend our basic model of a financial network with senior and junior debts. In our model only the central bank will be considered a senior creditor. The clearing vector has to be calculated accordingly. 
The recovery rate vector and payments satisfy the following equations, where the central bank is bank $0$.
\begin{align*}
    L_{u} & = \sum_{v \in N\setminus\{0\}} l_{u,v} \\
    r_u & = \min \left\{ \frac{a'_u}{L_u} , 1 \right\} \\
    p_{u,v} & = r_{u} l_{u,v} \text{ for all } u,v \in N\setminus\{0\}\\
    p_{u,0} & =
    \begin{cases}
        l_{u,0} & \text{if}\ a_u \geq L_u + l_{u,0} \\
        \min \{ \beta a_u, l_{u,0} \} & \text{if}\ a_u < L_u + l_{u,0}
    \end{cases} \\
    a_u & = x_u + \sum_{v \in N} p_{v,u} \\
    a'_u & =
    \begin{cases}
        a_u - l_{u,0} & \text{if}\ a_u \geq L_u + l_{u,0} \\
        \beta a_u - p_{u,0} & \text{if}\ a_u < L_u + l_{u,0}
    \end{cases}
\end{align*}
If bank $u$ defaults, we denote by $\delta_u$ the loss that the central bank endures from $u$'s default.
\begin{equation*}
    \delta_u = l_{u,0} - p_{u,0} = \max \left\{ l_{u,0} - \beta a_u, 0) \right\}
\end{equation*}
\end{toappendix}
\begin{definition}[Welfare Loss]
\label{def:welfare_loss}
    The \emph{welfare loss} is defined as
    \begin{equation*}
        WL = (1 - \beta) \sum_{u \in N^{-}} a_u + \lambda \left( \sum_{u \in N^{-}} \delta_u + B \right)
    \end{equation*}
    where $B$ is the total spend by the central bank on bailouts, $N^{-}$ is the set of insolvent banks, and all values are calculated after the central bank bailouts have been taken into account.
\end{definition}

The first term is the total default loss in the network. The second term is the loss the central bank incurs due to its claims not being paid in full plus the amount it spends on bailouts. The second term is weighted by $\lambda$, where $\lambda$ controls how important the central bank's own value is compared with interbank default losses. If $\lambda=0$, then this is equivalent to minimizing the total default loss, which is trivially solved by bailing out all banks.
In the limit as $\lambda \rightarrow \infty$ it is equivalent to the central bank maximizing its own market value.

\begin{figure}[H]
\centering
\begin{tikzpicture}[scale=0.6, every node/.style={scale=1.0}, >={Triangle[width=2mm,length=1mm]},->]
    \node[circle, draw, minimum size=0.5cm] (A) at  (0,2) {$u$};
    \node[circle, draw, minimum size=0.5cm] (B) at  (5,3) {$v$};
    \node[circle, draw, minimum size=0.5cm] (C) at  (5,0) {$w$};
    \node[circle, draw, minimum size=0.5cm] (D) at  (10,-1.5) {$s$};
    \node[circle, draw, minimum size=0.5cm] (0) at  (10,1.5) {$0$};
    \draw [semithick,->] (A) -- node[above] {\small $5$} (B);
    \draw [semithick,->] (A) -- node[above] {\small $5$} (C);
    \draw [line width=0.5mm,->] (B) -- node[above] {\small $10$} (0);
    \draw [line width=0.5mm,->] (C) -- node[above] {\small $10$} (0);
    
    \draw [semithick,->] (C) -- node[above] {\small $20$} (D);
    \node[rectangle, fill=white, draw, below right = -0.1cm of A] (a) {\footnotesize $8$};
    \node[rectangle, fill=white, draw, below right = -0.1cm of B] (b) {\footnotesize $6$};
    \node[rectangle, fill=white, draw, below right = -0.1cm of C] (c) {\footnotesize $20$};
    \node[rectangle, fill=white, draw, right = 0.0cm of D] (d) {\footnotesize $10$};
    \node[rectangle, fill=white, draw, below right = -0.1cm of 0] (g) {\footnotesize $0$};
\end{tikzpicture}
\caption{
Example of a financial network with a central bank (bank $0$) and senior debts (bold). The default cost parameter is $\beta = 1/2$. The central bank wants to minimize welfare loss with $\lambda = 2$. 
Initially all banks apart from $0$ and $s$ are insolvent and total assets are $\vec{a} := \{a_u, a_v, a_w, a_s, a_0\} = \{8, 8, 22, 11, 14\}$. So welfare loss is $WL = \frac{1}{2} \left( 8 + 8 + 22 \right) + 2 \left( 6 + 0 \right) = 31$. It is not worth bailing out bank $w$, as the welfare loss increases to $WL = 36$. However we see that it is worth bailing out bank $v$ at a cost of $\Delta_v = 2$, and it is even better to bail out bank $u$ (thereby saving $v$ indirectly). Once we bail out bank $u$, we can see that it is now also worth bailing out bank $w$ at an additional cost of $5$.
} 
\label{fig:welfare_example}
\end{figure}

Figure \ref{fig:welfare_example} is an example of a financial network with a central bank and senior debts and with the objective of minimizing the welfare loss. Bank $0$ is the central bank and all liabilities to the central bank are senior. The central bank wants to minimize welfare loss with parameter $\lambda = 2$. The optimal solution is to bail out banks $u$ and $w$ giving a welfare loss of $WL = 14$. Note however that bailing out only bank $w$ ($WL = 36$) is worse than not bailing out any banks at all ($WL = 31$).
We see that the optimal bailout strategy heavily depends on the network connections. One cannot view banks independently. In fact in the following theorem we show that minimizing the welfare loss is NP-hard.

\begin{theorem1rep}
\label{thm:welfare_loss}
Consider a financial network $(N, L, x)$. A central bank, bank $0$, wants to minimize the welfare loss, see Definition \ref{def:welfare_loss}, with $\lambda \geq 1$. All liabilities to the central bank are senior and all other liabilities are junior. For any default cost parameter $\beta<1$, deciding which banks to bail out to minimize welfare loss is NP-hard.
\end{theorem1rep}

\begin{proofsketch}
As in Theorem \ref{thm:single_unlimited}, we again give a reduction from \textsc{Vertex Cover}. However we have to introduce an extra gadget, a so-called ``black hole'' to make the proof work. Again we want to make sure that the vertex banks ``cover'' the edge banks, but this time the edge banks will not be solvent. The key intuition is that since central bank debts are senior, it is worth making sure the edge banks have enough incoming assets to pay off these prioritised central bank debts, but once these can be paid, any additional assets will just disappear down the black hole. Indeed any further reduction in default losses in the network will be almost completely lost. So we just want to ``cover'' the edge banks with the minimum number of vertex banks. The details are involved and are left for the appendix.
\end{proofsketch}

\begin{appendixproof}


We need an extra gadget for this proof, which we refer to as a black hole. The black hole consists of a string of $m$ banks, each with a high debt to the next and with $0$ external assets. At each node along the path a $(1-\beta)$ fraction of assets is lost such that only a $\beta^m$ fraction remains at the end. See Figure \ref{fig:black_hole}. 

\begin{figure}[H]
\centering
\begin{tikzpicture}[scale=0.6, every node/.style={scale=1.0}, >={Triangle[width=2mm,length=1mm]},->]
    \node (e1) at  (-3,0) {};
    \node[circle, draw, minimum size=0.9cm, label=center:\small $BH_0$] (BH_0) at  (0,0) {};
    \node[circle, draw, minimum size=0.9cm, label=center:\small $BH_1$] (BH_1) at  (4,0) {};
    \node[circle, draw, minimum size=0.9cm, label=center:\small $BH_2$] (BH_2) at  (8,0) {};
    \node (BH_i) at  (11.5,0) {...};
    \node[circle, draw, minimum size=0.9cm, label=center:\small $BH_m$] (BH_m) at  (15,0) {};
    \draw [semithick,->] (e1) -- node[above left] {\footnotesize $A$} (BH_0);
    \draw [semithick,->] (BH_0) -- node[above] {\footnotesize $X$} (BH_1);
    \draw [semithick,->] (BH_1) -- node[above] {\footnotesize $X$} (BH_2);
    \draw [semithick,->] (BH_2) -- node[above] {\footnotesize $X$} (BH_i);
    \draw [semithick,->] (BH_i) -- node[above] {\footnotesize $X$} (BH_m);
    \node[rectangle, fill=white, draw, below right = -0.1cm of BH_0] (bh0) {\footnotesize $0$};
    \node[rectangle, fill=white, draw, below right = -0.1cm of BH_1] (bh1) {\footnotesize $0$};
    \node[rectangle, fill=white, draw, below right = -0.1cm of BH_2] (bh2) {\footnotesize $0$};
    \node[rectangle, fill=white, draw, below right = -0.1cm of BH_m] (bhm) {\footnotesize $0$};
\end{tikzpicture}
\caption{Figure shows the construction of a black hole, where total assets of $BH_0$ are $A<X$. As long as the total assets of bank $BH_0$ are below $X$, then due to default costs the amount reaching $BH_m$ will only be $\beta^m A$.} 
\label{fig:black_hole}
\end{figure}

As in Theorem \ref{thm:single_unlimited}, we again give a reduction from \textsc{Vertex Cover}. We add such a black hole to our previous construction and choose $m$ and the weights carefully so that the reduction works. Given an instance of vertex cover, $(G=(V,E), k)$, we construct an instance of the bailout decision problem as follows:
\begin{itemize}
    \item We add bank $0$ with infinite external assets $x_0 = \infty$.
    \item Let $m =log_{\beta} \left( 1/4n^2 \right)$. We add a string of $m$ black hole banks, $\{BH_0, BH_1, ..., BH_m\}$, with external assets $0$ and liabilities $l_{BH_i, BH_{i+1}} = 6n^2$.
    \item We define $\epsilon = \frac{1}{2} \beta \left( 1 - \beta \right)$. For each $e \in E$ we add a bank $e$ with
    \begin{align*}
        x_e &= 0 \\
        l_{e, 0} &= \beta(1 + \beta) \\
        l_{e, BH_0} &= 5
    \end{align*}
    \item For each $v \in V$ we add a node $v$ with external assets $x_{v} = d_v$ and debts $l_{v, e} = 1 + \frac{1}{d_v}\epsilon$ for all edges $e$ that are incident to $v$.
    \item We want a welfare loss at least as low as $k' = (2 - \beta(1+\beta))\abs*{E} + k \lambda \epsilon)$
\end{itemize}
See Figure \ref{fig:bank_vertex_cover_welfare} for an illustration. 

\begin{figure}[H]
\centering
\begin{tikzpicture}[scale=0.6, every node/.style={scale=1.0}, >={Triangle[width=2mm,length=1mm]},->]
    \node[circle, draw, minimum size=0.75cm] (0) at  (14,-2) {$0$};
    \node[circle, draw, minimum size=0.75cm, label=center:\small $BH_0$] (BH0) at  (14,3) {};
    \node[minimum size=0.5cm, label=center:...] (BHi) at  (16,3) {};
    \node[circle, draw, minimum size=0.75cm, label=center:\small $BH_m$] (BHm) at  (18,3) {};
    \node[circle, draw, minimum size=0.5cm] (e1) at  (7,4) {$e_1$};
    \node[circle, draw, minimum size=0.5cm] (e2) at  (7,2) {$e_2$};
    \node[circle, draw, minimum size=0.5cm] (e3) at  (7,0) {$e_3$};
    \node (e4) at  (7,-3) {...};
    \node[circle, draw, minimum size=0.5cm] (v1) at  (0,4) {$v_1$};
    \node[circle, draw, minimum size=0.5cm] (v2) at  (0,0) {$v_2$};
    \node (v3) at  (0,-3) {...};
    \draw [line width=0.5mm,->] (e1) -- (0);
    \draw [line width=0.5mm,->] (e2) -- (0);
    \draw [line width=0.5mm,->] (e3) -- node[below] {\small $\beta(1+\beta)$} (0);
    \draw [line width=0.5mm,->] (e4) -- (0);
    \draw [semithick,->] (e1) -- node[above] {\small $5$} (BH0);
    \draw [semithick,->] (e2) -- node[above] {\small $5$} (BH0);
    \draw [semithick,->] (e3) -- node[above] {\small $5$} (BH0);
    \draw [semithick,->] (e4) -- (BH0);
    \draw [semithick,->] (BH0) -- node[above right] {\small $6n^2$} (BHi);
    \draw [semithick,->] (BHi) -- (BHm);
    \draw [semithick,->] (v1) -- node[above] {\small $1 + \frac{\epsilon}{d_{v_1}}$} (e1);
    \draw [semithick,->] (v1) -- node[above] {\small $1 + \frac{\epsilon}{d_{v_1}}$} (e2);
    \draw [semithick,->] (v1) -- node[above] {\small $1 + \frac{\epsilon}{d_{v_1}}$} (e3);
    \draw [semithick,->] (v2) -- node[below] {\small $1 + \frac{\epsilon}{d_{v_2}}$} (e2);
    \draw [semithick,->] (v2) -- node[above] {\small $1 + \frac{\epsilon}{d_{v_2}}$} (e4);
    \draw [semithick,->] (v3) -- (e4);
    \node[rectangle, fill=white, draw, below right = -0.1cm of 0] (d) {\footnotesize $n$};
    \node[rectangle, fill=white, draw, below right = -0.1cm of e1] (a) {\footnotesize $0$};
    \node[rectangle, fill=white, draw, below right = -0.1cm of e2] (a) {\footnotesize $0$};
    \node[rectangle, fill=white, draw, below right = -0.1cm of e3] (a) {\footnotesize $0$};
    \node[rectangle, fill=white, draw, below right = -0.1cm of v1] (b) {\footnotesize $d_{v_1}$};
    \node[rectangle, fill=white, draw, below right = -0.1cm of v2] (b) {\footnotesize $d_{v_2}$};
    \node[rectangle, fill=white, draw, below right = -0.1cm of BH0] (b) {\footnotesize $0$};
    \node[rectangle, fill=white, draw, below right = -0.1cm of BHm] (b) {\footnotesize $0$};
\end{tikzpicture}
\caption{Figure shows the construction used in Theorem \ref{thm:welfare_loss} to give a reduction from \textsc{Vertex Cover} to the single bank market value optimization problem.} 
\label{fig:bank_vertex_cover_welfare}
\end{figure}

Suppose there is a solution, $S$, to the vertex cover decision problem. We can bail out the banks corresponding to these vertices at a total cost of $k\epsilon$. Then each edge bank has at least $(1+\beta)$ incoming assets, so after deduction of default costs, it is still able to pay off the senior debt of $\beta (1+\beta)$ to bank $0$. Almost all of the rest of the assets in the network are lost through default down the black hole. Therefore the total welfare loss is $WL \leq \sum_{i} d_{v_i} - \beta(1+\beta)\abs*{E} + \lambda(0+k\epsilon) = k'$.

Now suppose there is a solution, $T$, to the welfare bailout decision problem. We show that there is another solution, $T'$, with only vertex banks, and at most $k$ vertex banks, that saves all edge banks. The vertices corresponding to the banks in $T'$ form a vertex cover in $G$ of size at most $k$.


Suppose there is a black hole bank in $T$. 
Even if $BH_0$ receives all of its claims in full, it still has total assets $a_{BH_0} = 5 \abs*{E} < 5n(n-1)/2 < 3n^2$, and total debt $6n^2$. So if the central bank bails out $BH_0$ (or any bank $BH_i$) then this costs $\Delta_{BH_0} > 3n^2$ and saves less than $(1-\beta)3n^2 < 3n^2$ in default losses. Moreover the central bank receives no assets back as there are no paths from any black hole bank to the central bank, bank $0$. Since $\lambda \geq 1$, 
the bailout cost outweighs the drop in default costs. So we can simply remove the black hole bank from $T$, without increasing $WL$.

Now suppose there is an edge bank in $T$. 
Let's consider edge bank $e=(u,v)$. Firstly, it is cheaper to first bail out its 2 corresponding vertex banks, $u$ and $v$, as these each have shortfall $\epsilon$, but add more than $1-\beta > \epsilon$ to bank $e$'s total assets. With $u$ and $v$ bailed out, bank $e$ has enough assets to pay off the senior liability to the central bank since $\beta a_e = \beta \left( 2 + \frac{\epsilon}{d_v} + \frac{\epsilon}{d_u} \right) > 2 \beta > \beta (1+\beta)$, so all further assets will go towards the black hole. So before bailout, bank $e$ will have total assets at most $a_e \leq 2 + 2\epsilon = 2 + \beta(1-\beta) < 2 + \beta$ and total debt $L_e = 5+\beta(1+\beta) > 5 + \beta$. Therefore it will have a shortfall of at least $\Delta_e > 5 + \beta - (2+\beta) = 3$, which is greater than the $(1-\beta)a_e < (1-\beta)(2+\beta) \leq 2$ that would be saved in default losses\footnote{Moreover the default loss savings would mostly even be lost along the black hole.}. So with $\lambda \geq 1$ it is not worth bailing out an edge bank. So we can replace any edge bank in $T$ with its corresponding vertex banks, without increasing $WL$.

Now suppose there is some edge bank $e=(u,v)$ that is not ``covered'', i.e., neither $u$ nor $v$ is in $T$.
Then bank $e$
has total assets $a_e = 2\beta$, $\beta$ incoming assets from each vertex bank $u$ and $v$. So after default costs, the central bank currently receives $2\beta^2$ from $e$ and is losing the remaining $\delta_e = \beta(1+\beta) - 2\beta^2 = \beta(1-\beta)$. Now if $u$ or $v$ is bailed out, wlog it is bank $v$, then $a_e = \beta + \left(1 + \frac{1}{d_v}\epsilon \right) > 1 + \beta$. Therefore after default costs are deducted, the central bank will still receive the $\beta(1+\beta)$ liability in full. Now the cost to bail out $v$ is $\epsilon$ and the gain in value for the central bank is $\beta(1-\beta) > \epsilon$. The central bank gains more than the cost of the bailout so the welfare loss decreases. 
So for any edge bank $e=(u,v)$ with $u$ and $v$ not in $T$, we can add one of the two vertex banks to $T$ without increasing $WL$.

So given a solution $T$ to the welfare bailout decision problem, we now have another solution, $T'$, with only vertex banks, such that all edge banks are ``covered'' and the $WL(T') \leq W(T) \leq k' = \sum_{i} d_{v_i} - \beta(1+\beta)\abs*{E} + \lambda(0+k\epsilon)$. Lastly we show that with this $WL$, the number of vertex banks $\abs*{T'} \leq k$.
We already have a vertex cover that ``covers'' the edge banks so losses from unpaid liabilities to bank $0$ are $\sum_{i \in N^{-}} \delta_i = 0$, where $N^{-}$ denotes the set of insolvent banks. 
So we are only interested in the default loss term and the total bailout cost $B$. 
For contradiction assume $\abs*{T'} > k$. Then default losses must be less than $\sum_{i} d_{v_i} - \beta(1+\beta)\abs*{E} - \lambda \epsilon$.
On the other hand 
But we know that any reduction in the default loss ends up as value directed towards the black hole, since all banks are in default and all edges (that are not already been paid in full) are directed towards the black hole. The total external assets of all the vertices is $\sum_{v \in G} d_v = 2 \abs*{E} \leq n(n-1) < n^2$. Currently a $(1-\beta)$-fraction of this is lost by the time it reaches the edge banks. So in the best case less than $(1-\beta)n^2$ in default losses can be saved before reaching the edge banks. Of these default loss savings a $\beta$-fraction will reach $BH_0$ and only a $\beta^{(m+1)}$-fraction will reach $BH_m$. The rest is lost through default. This leaves at most $(1-\beta)\beta^{(m+1)}n^2$ left in default loss savings. And
\begin{align*}
    (1-\beta)\beta^{(m+1)}n^2 
    &= (1-\beta)\beta \frac{1}{4n^2} n^2 \\
    &= \frac{1}{4}(1-\beta)\beta \\
    &= \frac{\epsilon}{2}
\end{align*}
Therefore the default loss term in the welfare loss can differ by at most $\epsilon/2$, which cannot compensate for the extra bailout cost (with $\lambda \geq 1$). 
Therefore $\abs*{T'} \leq k$, so the vertices corresponding to $T'$ form a vertex cover in $G$ of size at most $k$.

\end{appendixproof}

\begin{toappendix}
The key intuition is that since central bank debts are senior, it is worth making sure the edge banks have enough incoming assets to pay off their central bank debts, but once these can be paid off, any additional incoming assets will just disappear down the black hole. This means any additional reduction in default losses will be almost completely lost.
\end{toappendix}

\section{Abuse of the System}

We now consider another problem with bailouts. Suppose the central bank has a fixed, agreed-upon bailout strategy that is known to all participants in the system. For example the central bank might decide on a certain welfare function and always aim to maximize this in the event of a crisis. This makes the central bank's actions predictable. Thus, in the anticipation of a crisis, banks can manipulate the network 
to their benefit.
A naive strategy might be to remove liabilities to insolvent banks in order to increase one's value. However, in this case there is no incentive for the insolvent bank to agree. We show an example of 2 banks making a new debt contract that strictly benefits them both in the event of a bailout. 

\begin{figure}[H]
\centering
    \begin{subfigure}[t]{.45\textwidth}
    \centering
        \begin{tikzpicture}[scale=0.6, every node/.style={scale=1.0}, >={Triangle[width=2mm,length=1mm]},->]
            \node[circle, draw, minimum size=0.5cm] (A) at  (0,0) {$u$};
            \node[circle, draw, minimum size=0.5cm] (B) at  (3,0) {$v$};
            \node[circle, draw, minimum size=0.5cm] (C) at  (6,2) {$w$};
            \node[circle, draw, minimum size=0.5cm] (O) at  (6,-2) {$0$};
            \draw [semithick,->] (A) -- node[above] {$4$} (B);
            \draw [line width=0.5mm,->] (B) -- node[below] {$2$} (O);
            \draw [line width=0.5mm,->] (C) -- node[right] {$2$} (O);
            \node[rectangle, fill=white, draw, below right = -0.1cm of A] {\footnotesize $2$};
            \node[rectangle, fill=white, draw, below right = -0.1cm of B] {\footnotesize $0$};
            \node[rectangle, fill=white, draw, below right = -0.1cm of C] {\footnotesize $2$};
            \node[rectangle, fill=white, draw, below right = -0.1cm of O] {\footnotesize $0$};
        \end{tikzpicture}
    \caption{Initial state of the financial network}
    \label{fig:bailout_abuse_a}
    \end{subfigure}%
    \qquad
    \begin{subfigure}[t]{.45\textwidth}
    \centering
        \begin{tikzpicture}[scale=0.6, every node/.style={scale=1.0}, >={Triangle[width=2mm,length=1mm]},->]
            \node[circle, draw, minimum size=0.5cm] (A) at  (0,0) {$u$};
            \node[circle, draw, minimum size=0.5cm] (B) at  (3,0) {$v$};
            \node[circle, draw, minimum size=0.5cm] (C) at  (6,2) {$w$};
            \node[circle, draw, minimum size=0.5cm] (O) at  (6,-2) {$0$};
            \draw [semithick,->] (A) -- node[above] {$4$} (B);
            \draw [line width=0.5mm,->] (B) -- node[below] {$2$} (O);
            \draw [line width=0.5mm,->] (C) -- node[right] {$2$} (O);
            \draw [semithick,->,red] (B) -- node[above] {$2$} (C);
            \node[rectangle, fill=white, draw, below right = -0.1cm of A] {\footnotesize $2$};
            \node[rectangle, fill=white, draw, below right = -0.1cm of B] {\footnotesize \textcolor{red}{$1$}};
            \node[rectangle, fill=white, draw, below right = -0.1cm of C] {\footnotesize \textcolor{red}{$1$}};
            \node[rectangle, fill=white, draw, below right = -0.1cm of O] {\footnotesize $0$};
        \end{tikzpicture}
    \caption{State of the network after banks $v$ and $w$ have agreed a debt contract, whereby $w$ lends $1$ to $v$ in return for a debt of $2$}
    \label{fig:bailout_abuse_b}
    \end{subfigure}%
\caption{Figure shows an example of how 2 banks could abuse the bailout strategy of the central bank in order to mutually increase their final market values (at the expense of the central bank). The default cost parameter is $\beta= 1/2$ and the welfare loss parameter is some $\lambda > 1$.}
\label{fig:bailout_abuse}
\end{figure}

Consider the financial network in Figure \ref{fig:bailout_abuse}. The initial state of the network is as shown in Figure \ref{fig:bailout_abuse_a}. Bank $0$ is the central bank, debts to the central bank are senior (prioritised) and the central bank wants to minimize the welfare loss, as defined in Definition \ref{def:welfare_loss}, with $\lambda > 1$. The default cost parameter is $\beta = 1/2$. 
Let $\mathcal{D}$ denote the set of insolvent banks that bank $0$ decides to bail out.
Initially banks $u$ and $v$ are insolvent and $w$ is not, but if bank $0$ bails out $u$, then $v$ is also solvent so there are really only $3$ options: $\mathcal{D} \in \left\{ \varnothing, \{u\}, \{v\} \right\}$. You can see the welfare loss in each case in Table \ref{fig:table_abuse_a}. 
The best strategy is to bail out bank $v$ directly. This results in market value of $0$ for all banks $u$, $v$ and $w$. 
On the other hand, after a new debt contract, whereby $w$ gives $1$ to $v$ in return for a debt of $2$ (as illustrated in Figure \ref{fig:bailout_abuse_b}), $u$, $v$ and $w$ are all insolvent. By the same ``chain effect'' logic as before, the central bank has 4 options on whom to bail out $\mathcal{D} \in \left\{ \varnothing, \{u\}, \{v\}, \{w\} \right\}$. You can see the welfare loss for each case in Table \ref{fig:table_abuse_b}. 
The best strategy now is to bail out bank $u$. And the market value for banks $v$ and $w$ in this case is now $1$, whilst the market value for the central bank has dropped to $2$. In this way banks $v$ and $w$ have increased their market value after bailout by arranging a new debt contract.

\begin{figure}[H]
\centering
    \begin{subfigure}[t]{\textwidth}
    \centering
        \begin{adjustbox}{max width=\textwidth}
        \begin{tabular}{|c|c|c|}
            \hline
            $\mathcal{D}$ & Welfare Loss & Market Value $(u,v,w,0)$ \\
            \hline
            $\varnothing$ & $0.5(2 + 1) + \lambda(1.5+0) = 1.5 + 1.5\lambda$ & $(0,0,0,2.5)$ \\
            $\{u\}$ & $0.5(0) + \lambda(0+2) = 2\lambda$ & $(0,2,0,2)$ \\
            $\bm{\{v\}}$ & $\bm{0.5(2) + \lambda(0+1) = 1 + \lambda}$ & $\bm{(0,0,0,3)}$ \\
            \hline
        \end{tabular}
        \end{adjustbox}
        \caption{Welfare Loss for all the bailout options of the central bank in Figure \ref{fig:bailout_abuse_a}. Given $\lambda > 1$, we see that welfare loss is lowest when the central bank bails out bank $v$ only.}
        \label{fig:table_abuse_a}
    \end{subfigure}%
    \vspace*{10pt}
    \begin{subfigure}[t]{\textwidth}
    \centering
        \begin{adjustbox}{max width=\textwidth}
        \begin{tabular}{|c|c|c|}
            \hline
            $\mathcal{D}$ & Welfare Loss & Market Value $(u,v,w,0)$ \\
            \hline
            $\varnothing$ & $0.5(2 + 2 + 1) + \lambda(2.5) = 2.5 + 2.5\lambda$ & $(0,0,0,1.5)$ \\
            $\bm{\{u\}}$ & $\bm{0.5(0) + \lambda(0+2) = 2\lambda}$ & $\bm{(0,1,1,2)}$ \\
            $\{v\}$ & $0.5(2) + \lambda(0+2) = 1 + 2\lambda$ & $(\bm{0},0,1,2)$ \\
            $\{w\}$ & $0.5(2 + 2) + \lambda(1+1) = 2 + 2\lambda$ & $(0,0,0,2)$ \\
            \hline
        \end{tabular}
        \end{adjustbox}
        \caption{Welfare Loss for all the bailout options of the central bank in Figure \ref{fig:bailout_abuse_b}. Given $\lambda > 1$, we see that welfare loss is lowest when the central bank bails out bank $u$.}
        \label{fig:table_abuse_b}
    \end{subfigure}%
    \caption{Welfare Loss for all bailout options of the central bank before and after the new contract between $v$ and $w$. We see that the argmin changes from bailing out $\{v\}$ to bailing out $\{u\}$.}
    \label{fig:table_abuse}
\end{figure}

The intuition behind the example is as follows. The central bank wants to bail out bank $v$ so that bank $v$ can pay its liability $l_{v,0} = 2$ in full. Initially it is cheaper to bail out bank $v$ directly, but bank $v$ would prefer that bank $u$ be bailed out instead. To force the central bank into doing so, bank $v$ takes on more debt to make a direct bailout more expensive. It can for example do this by borrowing from bank $w$, who in return gets a cut of the gains in the form of interest on the loan, i.e. bank $v$ will pay back more than it borrows. This is all at the eventual expense of the central bank, who now chooses to bail out bank $u$.


\section{Conclusion}

In this paper we introduce several models of financial networks and focus on the question of which banks to bail out in a financial crisis. 
We formalize this question and consider various natural objective functions that a central bank may wish to optimize.

We show that maximizing a bank's market value is NP-hard as is minimizing welfare loss, where welfare loss balances the central bank's value with the default losses in the network. 
These two results are particularly relevant for central banks.
The central bank can be interpreted as a society or government node and in a crisis a central bank will aim to minimize government costs and impact on society, whilst also saving 
businesses from bankruptcy. We prove that balancing these objectives gives a NP-hard optimization problem.
These are the first results proving that finding the optimal bailout strategy in a crisis is not only practically complex, but provably NP-hard to do. As a next step one should look for approximation algorithms. We leave this for future work.

Furthermore we show in this paper that a transparent bailout system can be abused. Future research should look at the game theoretic implications and at defining bailout objectives that do not incentivize such behavior.

\newpage
\printbibliography
\end{document}